\documentclass[twocolumn,showpacs, pre, floatfix]{revtex4}

\usepackage{epsfig}
\usepackage{bm}

\begin{document}

\title{Deriving nonequilibrium interface kinetics from variational principles}
\author{Robert Spatschek}
\email{r.spatschek@fz-juelich.de}
\affiliation{Institut f\"ur Festk\"orperforschung, Forschungszentrum J\"ulich, D-52425 J\"ulich, Germany}
\date{\today}
 
\begin{abstract}
Nonequilibrium dynamics at interfaces is generally driven by a chemical potential.
Here we demonstrate a generic technique to derive the basic equations of motion, boundary conditions and the chemical potential in a consistent way from fundamental variational principles.
As a particular example, we consider a solid surface with elastodynamic effects, together with surface energy and tension.
We apply the generic results to perform a linear stability analysis of a planar front subjected to uniaxial stress (Asaro-Tiller-Grinfeld instability \cite{ATG}), here also with surface tension.
\end{abstract}


\maketitle

\section{Introduction}

The quest for the correct equations of motion describing a system is an essential task of each problem in all branches of physics.
The procedure is well-known for mechanical systems, where for example the appropriate Lagrangian has to be found, from which the (Newtonian) equations of motion can be derived using variational principles.
This method can in particular be applied to continuum mechanics, generating the usual bulk equations and boundary conditions.

However, for more complicated dissipative systems, these straightforward and general methods are usually not applicable.
Bulk diffusion, friction, and also irreversible phase transitions are important examples.
In the framework of linear nonequilibrium thermodynamics, the ``process velocity'' is proportional to the deviation of an energy functional from its equilibrium value.

If we consider for example diffusion processes, then the flux of material is typically $j=-D\nabla\mu$ with the diffusion coefficient D and a chemical potential $\mu$.
Similarly, for phase transformations at an interface, the resulting interface normal velocity is assumed to be proportional to the difference in chemical potentials in both phases, $v_{n}=\kappa \Delta\mu$ ($\kappa$ is a kinetic coefficient);
obviously, in both cases the chemical potential $\mu$ is the relevant driving force for this type of dynamics;
however, it is not evident how it can be expressed through the state variables of the system. 
Therefore it is desirable to provide a method to derive the correct expression for the chemical potential using first principle techniques.
For relatively simple processes, it is easy to guess it correctly, but as soon as more complicated effects are included, the results are sometimes unexpected.
A particular example are elastodynamic effects, and the correct expression for the chemical potential will be derived in this paper.
The influence of surface tension, also causing complicated contributions to the chemical potential, was already considered for processes at elastically stressed surfaces in \cite{Grilhe93};
however, the author made some systematic mistakes.
This illustrates that the lack of a consistent derivation and the use of bare intuition can easily lead to wrong predictions.
Though the aim of the current paper is mainly to demonstrate methodological aspects how to derive consistently these expressions from variational principles, we also want to point out that the results form a basis to the understanding of important topics like solidification of superfluid helium \cite{Uwaha86} and fracture mechanics \cite{Brener03}.

The Asaro-Tiller-Grinfeld (ATG) instability is a particular example of these processes \cite{ATG}.
Here one considers a solid piece of material which is either in contact with its melt such that phase transitions and thus motion of the interface are possible, or which suffers from surface diffusion.
A uniaxial stress at the surface of this solid gives rise to a morphological instability;
long wave deformations lead to a reduction of the total energy.
A relaxation of elastic stresses in some regions makes them more favorable in comparison to the valleys in which the stress is increased;
consequently, a corrugated surface is characterized by a modulation of the elastic energy density at the surface.
This contribution favors a destabilization of the interface, counteracted by the surface energy which prefers a flat surface.
After the initial development, the originally slow process becomes extremely fast and leads to an unphysical breakdown of the analytical and numerical predictions:
Deep groves with sharp tips advancing with increasing velocity can form in the sample \cite{Kassner99,Chiu93}.
Here strong surface tension and also dynamical effects become relevant, but they are not taken into account in conventional descriptions.
The main problem is here that it is not obvious how these effects enter into the expression of the chemical potential.

The main goal of the current paper is therefore to provide a method to derive all describing quantities in a consistent way from variational principles.
A particular illustration is the Grinfeld instability with surface tension.
The paper is organized as follows:
In Section \ref{Lagrangian} we present the Lagrangian which is used to derive all further physical laws.
In particular, we consider a linear elastic medium with dynamical effects, surface tension and surface energy, under the influence of external forces.
In Section \ref{Elastodynamics} we derive the basic equations of motion and boundary conditions for the ``mechanical'' degrees of freedom, the elastic displacements, from the Lagrangian.
In Section \ref{chempot} the expression for the chemical potential at the surface of a solid body is derived using variational principles.
The surface tension enters into the final expression in a complicated way, and also inertial effects appear unexpectedly.
Therefore, in Section \ref{dissipation} the rate of change of the total energy during a reshuffling of the surface is calculated;
this confirms the preceding results for the chemical potential.
A specific application of these general techniques is the Grinfeld instability with surface tension (Section \ref{Grinfeld}).
We derive the spectrum of the initial stage of this instability.
Additionally, in Section \ref{Grilhe} the energetic change during the morphological transformations is calculated, correcting inconsistencies in \cite{Grilhe93}.
Appendix \ref{conservation} explains the fundamental conservation laws for momentum and angular momentum for elastic systems with surface tension;
Appendix \ref{deriv} contains the mathematical details to express the dissipation rate and the functional derivative for a generic functional which is used to derive the chemical potential and the energy dissipation.

\section{Action and Lagrangian}
\label{Lagrangian}

We consider a solid body in vacuum under the influence of external forces, linear elastodynamics, surface energy and tension.
The Lagrangian is given by
\begin{equation} \label{Lag::eq1}
L = \int\limits_{V(t)} (T-U_{b})\, dV - \int\limits_{\partial V(t)} U_{s}\,d\tau = L_{b} + L_{s},
\end{equation}
where $V(t)$ is the time-dependent volume of the elastic body and $\partial V(t)$ its boundary with elements $d\tau$.
The expressions for kinetic and potential energy densities are
\begin{eqnarray}
T &=& \frac{1}{2} \rho \dot{u}_{i} \dot{u}_{i}, \label{Lag::eq2} \\
U_{b} &=& \frac{1}{2} \sigma_{ik} u_{ik}, \label{Lag::eq3} \\
U_{s} &=& \alpha + \beta u_{\tau\tau} - \sigma_{ni}^{(0)}u_{i}. \label{Lag::eq4}
\end{eqnarray}
Here $\rho$ denotes the mass density, which we assume to be constant.
Therefore we neglect compressive effects, since $\Delta\rho=\rho u_{ii}\ll \rho$.
The elastic displacement field is denoted by $u_{i}$, $\sigma_{ik}$ is the stress and $u_{ik}$ the strain tensor, $\alpha$ is the surface energy density and $\beta$ the surface tension coefficient.
The representation of the surface tension contribution is unusual, and we will further comment on it below \cite{Marchenko80}.
The stresses $\sigma_{ij}^{(0)}$ are the given external forces which may act on some parts of the boundary; everywhere else they are zero. 
We assume the volume to be two-dimensional; $\tau$ is the tangential and $n$ the normal direction at the interface.

The surface {\em energy} in Eq. (\ref{Lag::eq4}) expresses the energy of the original undeformed surface, whereas the surface {\em tension} contribution gives the additional energy which arises from elongation of the surface due to elastic effects.
Notice that $u_{\tau\tau}d\tau$ is the elongation or compression of the length element $d\tau$.
Surface energy gives only an additional contribution to the total energy, surface tension also couples to the elastic fields and changes the elastic boundary conditions, as will be seen in the next section.
Also, in Section \ref{Grilhe}, it will be shown how the surface tension can be interpreted in a more intuitive way, by relating it to surface forces which are proportional to the local curvature of the boundary.
In general, both coefficients $\alpha$ and $\beta$ are of the same order of magnitude, and for a liquid they can be shown to be identical \cite{Gibbs28};
in general, $\beta$ can even be negative.
For three-dimensional situations, the surface tension contribution in Eq.~(\ref{Lag::eq4}) reads in general $U_{\beta} = \beta_{ij} u_{ij}$ with the summation of $i, j$ being restricted to the two tangential directions of the surface.
In the case of surfaces with a symmetry axis higher than twofold, the surface tension tensor is diagonal, $\beta_{ij} = \beta \delta_{ij}$ \cite{Marchenko80}.

Stress and strain are connected by Hooke's law.
For simplicity, we restrict it to the isotropic case,
\begin{equation} \label{Lag::eq5}
\sigma_{ij} = \frac{E}{1+\nu} \left(u_{ij} + \frac{\nu}{1-2 \nu} \delta_{ij} u_{kk}\right),
\end{equation}
with $E$ being the elastic modulus and $\nu$ the Poisson ratio.
The inverse transformation is
\begin{equation} \label{Lag::eq6}
u_{ij} = \frac{1}{E} \left[ (1+\nu)\sigma_{ij} - \nu\delta_{ij}\sigma_{kk} \right].
\end{equation}
The action is given by the time integral
\begin{equation} \label{Lag::eq7}
S = \int\limits_{t_{0}}^{t_{1}} L\, dt = S_{b} + S_{s}
\end{equation}
with arbitrary initial and end times $t_{0}, t_{1}$.

Since the Lagrangian is translational and rotational invariant in the absence of external forces, total momentum and angular momentum are conserved.
This property is not obvious if nonequilibrium processes at the surface locally change the amount of material;
a further discussion is given in Appendix \ref{conservation}.

\section{Elastodynamic equations and boundary conditions}
\label{Elastodynamics}

The Euler-Lagrange conditions demand that the variation of the action with respect to the displacements has to vanish \cite{Landau::elasticity}.
Here we also take into account that the volume can change in time due to other processes.
We obtain for the bulk contribution
\begin{eqnarray}
\delta S_{b} &=& \int\limits_{t_{0}}^{t_{1}} dt \int\limits_{V(t)} (\rho \dot{u}_i \delta
 \dot{u}_i - \sigma_{ij}\delta u_{ij}) dV \nonumber \\
&=& \int\limits_{t_{0}}^{t_{1}} dt \Big\{ \int\limits_{V(t)} ( \rho \dot{u}_i\delta \dot{u}_i
 + \frac{\partial \sigma_{ij}}{\partial x_j} \delta u_i) dV \nonumber \\
 && - \int\limits_{\partial V(t)} \sigma_{ni}\delta u_i d\tau \Big\}. \nonumber
\end{eqnarray}
Now we examine the first term
\[
\delta A = \int\limits_{t_{0}}^{t_{1}} dt \int\limits_{V(t)} \rho \dot{u}_i \delta \dot{u}_i dV
\]
in more detail.
We start from the expression
\begin{eqnarray}
\frac{d}{dt} \int\limits_{V(t)}
\rho \dot{u}_i \delta u_i dV &=& \int\limits_{V(t)}
 ( \rho \ddot{u}_i \delta u_i + \rho \dot{u}_i
 \delta\dot{u}_i ) dV \nonumber \\
&& + \int\limits_{\partial V(t)} v_n \rho \dot{u}_i \delta u_i d\tau, \label{mech::eq5} 
\end{eqnarray}
where $v_n$ is the normal velocity of the interface;
it is positive if the solid extends.
We point out that the elastic velocity $\dot{u}_n$ at the surface and $v_n$ are different quantities:
The latter one describes the morphological changes of the shape due to nonelastic effects, for example by surface diffusion or the growth of a solid body from its undercooled melt.
On the other hand, the normal displacement rate $\dot{u}_n$ does not contribute to $v_n$, since in the framework of the linear theory of elasticity one does not distinguish between the undeformed and deformed frame of reference.
Therefore we obtain
\begin{eqnarray*}
\delta A &=& \int\limits_{t_{0}}^{t_{1}} dt \Bigg\{ \frac{d}{dt} \int\limits_{V(t)}
 \rho \dot{u}_i \delta u_i dV - \int\limits_{V(t)} \rho
 \ddot{u}_i \delta u_i dV \\
&& - \int\limits_{\partial V(t)} v_n \rho \dot{u}_i \delta u_i d\tau  \Bigg\}.
\end{eqnarray*}
The first bulk integral does not contribute since we have to demand that all variations disappear at $t_{0}$ and $t_{1}$.
Consequently, we obtain the variation of the bulk action
\begin{eqnarray*}
\delta S_{b} = \int\limits_{t_{0}}^{t_{1}} dt &\Bigg\{ &\int\limits_{V(t)} [-\rho \ddot{u}_i
 + \frac{\partial \sigma_{ij}}{\partial x_j}] \delta u_i\, dV \\
& - & \int\limits_{\partial V(t)} [ \sigma_{ni} + v_n \rho \dot{u}_i] \delta u_i\, d\tau \Bigg\}.
\end{eqnarray*}
We define the curvature of the surface to be negative for a convex body.
The tangential strain is
\begin{equation} \label{mech::eq1}
u_{\tau\tau} = \frac{\partial u_{\tau}}{\partial\tau} = \frac{d u_{\tau}}{d \tau} - u_{n} \kappa.
\end{equation}
Here, the definitions of the partial and the total tangential derivative are
\begin{eqnarray*}
\frac{\partial u_{\tau}}{\partial \tau}  &=& \lim_{\Delta \tau\to 0} \frac{u_{\tau}(\tau+\Delta \tau) - u_{\tau}(\tau)}{\Delta\tau}, \\
\frac{d u_{\tau}}{d \tau}  &=& \lim_{\Delta \tau\to 0} \frac{u_{\tau+\Delta\tau}(\tau+\Delta \tau) - u_{\tau}(\tau)}{\Delta\tau}.
\end{eqnarray*}
Thus we get for the variation of the boundary action
\[
\delta S_{s} = \int\limits_{t_{0}}^{t_{1}} dt \int\limits_{\partial V(t)} \left((\beta\kappa+\sigma_{nn}^{(0)})\delta u_{n} +\sigma_{n\tau}^{(0)}\delta u_{\tau} \right) d\tau,
\]
since the term containing $d\delta u_{\tau}/d\tau$ can be integrated along the boundary and does not contribute.
Vanishing total variation, $\delta S_{b}+\delta S_{s}=0$, requires therefore
\begin{eqnarray}
\frac{\partial\sigma_{ij}}{\partial x_{j}} &=& \rho \ddot{u}_{i}, \label{mech::eq2} \\
\sigma_{nn} &=& -v_{n}\rho \dot{u}_{n} + \beta\kappa + \sigma_{nn}^{(0)}, \label{mech::eq3} \\
\sigma_{n\tau} &=& -v_{n}\rho \dot{u}_{\tau} + \sigma_{n\tau}^{(0)}. \label{mech::eq4}
\end{eqnarray}
These are the known elastodynamic equations, together with the boundary conditions, now also considering surface tension.
The velocity dependent term in the boundary conditions ensures momentum conservation.
Notice that we had chosen for simplicity the particular case of a solid body inside the vacuum, and not for example in contact with a liquid or different solid phase.
However, the calculations can easily be extended to these slightly more complicated case;
then in Eqs.~(\ref{mech::eq3}) and (\ref{mech::eq4}) the difference $[\sigma_{ni} + v_n\rho \dot{u}_i]$ between the values in the two phases appears instead \cite{Freund98}.

\section{The chemical potential}
\label{chempot}

In the previous section the dynamics of the mechanical variables, the elastic displacements, was studied, requiring that the variation of $S$ with respect to $u_i$ vanishes.
The elastic body was already considered as moving or transforming, which leads to the appearance of the interface normal velocity in the elastic boundary conditions.

The reason for this surface rearrangement is not yet specified and is not of mechanical origin.
A specific example would be surface diffusion.
The elastic motion as ``mechanical'' process describes the motion of individual atoms due to Newton's law;
in contrast, the ``chemical'' processes are related to a (local) change of the number of atoms and consequently of the shape of the body.

Demanding that the solid body is in equilibrium not only with respect to variations of the displacement vector but also with respect to this new degree of freedom, the interface position, leads to the requirement that $\delta S=0$ holds also for variations of the surface position in normal direction by an interface shift $\delta n$.
Otherwise, small deviations from the condition $\delta S=0$ initiate thermodynamical processes, driven by the chemical potential
\begin{equation} \label{chem::eq1}
\mu = -\Omega \frac{\delta L}{\delta n}.
\end{equation}
From the bulk Lagrangian we gain
\[
\mu_{b} = -\Omega \frac{\delta L_{b}}{\delta n} = \Omega (U_{b}-T).
\]
Notice that this expression contains in particular the {\em difference} between potential and kinetic bulk energy.
A naive guess would be that the chemical potential is the total energy density, containing thus the {\em sum} of both contributions.
However, the above explanation shows that this is not true.
A further discussion on this point is given in the next section.

For the boundary contribution we assume here that the surface of consideration is free of external tractions.
Then we get from Eq.~(\ref{deriv::eq1})
\begin{eqnarray*}
\mu_{s} &=& -\Omega \frac{\delta L_{s}}{\delta n} \\
&=& \Omega\, \left[ -\alpha\kappa + \beta \left( \frac{\partial u_{\tau\tau}}{\partial n} - u_{\tau\tau}\kappa 
- 2 \frac{du_{n\tau}}{d\tau} \right) \right]
\end{eqnarray*}
where we used the rotation rule
\begin{equation} \label{chem::eq4}
\frac{\partial u_{\tau\tau}}{\partial \theta} = 2 u_{n\tau}
\end{equation}
and the expression for the total derivative
\begin{equation} \label{chem::eq5}
\frac{d}{d\tau} = \frac{\partial}{\partial \tau} + \kappa \frac{\partial}{\partial\theta}.
\end{equation}
Therefore the total chemical potential, $\mu=\mu_b + \mu_s$, reads
\begin{eqnarray} 
\mu &=& \Omega \Bigg[ \frac{1}{2}\sigma_{ik}u_{ik} - \frac{1}{2}\rho \dot{u}_{i}^{2} -\alpha\kappa + \beta \Bigg( \frac{\partial u_{\tau\tau}}{\partial n} \nonumber \\
&& - u_{\tau\tau}\kappa - 2 \frac{du_{n\tau}}{d\tau} \Bigg) \Bigg]. \label{chem::eq6}
\end{eqnarray}
Obviously, one immediately reproduces the known elastostatic case with surface energy, without inertial and surface tension effects \cite{ATG};
notice that the chemical potential is then nothing else than the energy density at the surface.
Inertial effects enter in an unexpected way, because instead of the {\em sum} of elastic and kinetic energy density their {\em difference} appears here.
Surface tension gives rise to three contributions:
At first, the term $\partial u_{\tau\tau}/\partial n$ reflects the change of the strain $u_{\tau\tau}$ when the solid extends in normal direction, and $u_{\tau\tau}\kappa$ expresses the elongation of a curved surface element.
Without dynamical effects and external forces, the last term in the surface tension contribution vanishes, because, by virtue of the boundary condition (\ref{mech::eq4}) and Hooke's law, also $u_{n\tau}=0$.
This full derivative appears only if both inertial contributions and surface tension are taken into account, leading to a coupling of both effects.

\section{The dissipation rate}
\label{dissipation}

An alternative approach to derive the expression for the chemical potential is to express the dissipation of energy due to morphological changes of the solid phase in terms of boundary integrals.
Then the chemical potential gives the amount of energy change at the traction free surfaces (per atom) if the shape of the solid is altered.
It consists basically of three contributions:
A change of the stressed volume reduces or increases the energy by the density $\Omega(U_b+T)$;
also, a change of the shape modifies the boundary conditions and leads to an additional energy contribution $-2\Omega T$.
Finally, complicated ``boundary'' terms arising from surface tension and energy appear. 

To calculate the dissipation rate, we start with the bulk contribution,
\begin{equation} \label{diss::eq1}
E_{b} = \int\limits_{V(t)} (T+U_{b}) dV.
\end{equation}
Then we obtain
\begin{eqnarray*}
\frac{dE_{b}}{dt} &=& \int\limits_{\partial V(t)} v_{n}(T+U_{b})\, d\tau + \int\limits_{V(t)} \frac{\partial}{\partial t} (T+U_{b}) dV \\
&=& \int\limits_{\partial V(t)} v_{n}(T+U_{b})\, d\tau + \int\limits_{V(t)} (\rho\dot{u}_{i}\ddot{u}_{i} + \sigma_{ij} \dot{u}_{ij})\, dV \\
&=& \int\limits_{\partial V(t)} v_{n}(T+U_{b})\, d\tau + \int\limits_{V(t)} \big( \dot{u}_{i}\frac{\partial\sigma_{ij}}{\partial x_{j}} \\
&& + \frac{\partial}{\partial x_{j}}(\sigma_{ij}\dot{u}_{i}) - \dot{u}_{i} \frac{\partial\sigma_{ij}}{\partial x_{j}} \big)\, dV \\
&=& \int\limits_{\partial V(t)} v_{n}(T+U_{b})\, d\tau + \int\limits_{\partial V(t)} \sigma_{in}\dot{u}_{i}\, d\tau \\
&=& \int\limits_{\partial V(t)} v_{n}(T+U_{b})\, d\tau + \int\limits_{\partial V(t)} \big( (-v_{n}\rho \dot{u}_{n} + \beta\kappa) \dot{u}_{n}\\
&&  - v_{n}\rho\dot{u}_{\tau}^{2} \big)\, d\tau \\
&=& \int\limits_{\partial V(t)} v_{n}(U_{b}-T)\, d\tau + \beta\int\limits_{\partial V(t)} \kappa \dot{u}_{n}\, d\tau.
\end{eqnarray*}
The boundary energy is
\begin{equation} \label{diss::eq2}
E_{s} = \int\limits_{\partial V(t)} \left( \alpha + \beta u_{\tau\tau} - \sigma_{ni}^{(0)} u_{i} \right) d\tau,
\end{equation}
and its dissipation rate according to Eq.~(\ref{deriv::eq2})
\begin{eqnarray*}
\frac{dE_{s}}{dt} &=& \int\limits_{\partial V(t)} v_{n} \Bigg[ -\alpha\kappa + \beta \Bigg\{ \frac{\partial u_{\tau\tau}}{\partial n} - u_{\tau\tau}\kappa
- \frac{\partial u_{\tau\tau}}{\partial \tau\partial\theta}\\
&& - \kappa \frac{\partial^{2} u_{\tau\tau}}{\partial \theta^{2}} \Bigg\} \Bigg]\, d\tau
+ \int\limits_{\partial V(t)} \left[ \beta \dot{u}_{\tau\tau} + \sigma_{ni}^{(0)}\dot{u_{i}} \right]\, d\tau.
\end{eqnarray*}
Here the first integral is the contribution due to the change of the shape and {\em constant} fields, whereas the second one does the opposite, {\em fixed} shape and changing fields.
We assume explicitly that the external forces are constant in time and that $v_{n}=0$ on the interface parts where these forces act.
From (\ref{mech::eq1}) follows for fixed shape (thus $\dot{\kappa}=0$) in the second integral
\[
\dot{u}_{\tau\tau} = \frac{d\dot{u}_{\tau}}{d\tau} - \dot{u}_{n}\kappa.
\]
Using again Eqs.~(\ref{chem::eq4}) and (\ref{chem::eq5}), we do not get a contribution from the full derivative, since it can be integrated along the whole (closed) surface.
Thus we have
\begin{eqnarray*}
\frac{dE_{s}}{dt} &=& \int\limits_{\partial V(t)} \Bigg\{ v_{n} \Bigg[ -\alpha\kappa + \beta \Bigg(\frac{\partial u_{\tau\tau}}{\partial n}  - u_{\tau\tau}\kappa \\
&&  - 2\frac{du_{n\tau}}{d\tau} \Bigg) \Bigg]
+ \left(  -\beta\kappa \dot{u}_{n} + \sigma_{ni}^{(0)}\dot{u}_{i} \right) \Bigg\} d\tau.
\end{eqnarray*}
Hence the total change of energy, $E=E_b+E_s$, is
\begin{eqnarray*}
\frac{dE}{dt} &=& \int\limits_{\partial V(t)} v_{n} \Bigg[ U_{b} - T -\alpha\kappa + \beta \Bigg( \frac{\partial u_{\tau\tau}}{\partial n} - u_{\tau\tau}\kappa  \\
&& - 2\frac{du_{n\tau}}{d\tau} \Bigg) \Bigg] + \int\limits_{\partial V(t)} \sigma_{ni}^{(0)} \dot{u}_{i}\, d\tau.
\end{eqnarray*}
The last integral gives the trivial contribution resulting from the external forces.
We are mainly interested in local deformations of traction free surfaces ($\sigma_{ni}^{(0)}=0$), leading to an alternative definition of the chemical potential,
\begin{equation} \label{diss::eq3}
\frac{dE}{dt} = \frac{1}{\Omega} \int\limits_{\partial V(t)} \mu v_{n}\, d\tau
\end{equation}
with the same expression (\ref{chem::eq6}).
In particular, we clearly see that the density of the bulk {\em Lagrangian} instead of the total {\em energy} is relevant for the dynamics.

\section{Grinfeld instability with surface tension}
\label{Grinfeld}

A straightforward application of the preceding results is the Grinfeld instability \cite{ATG};
similar to \cite{Grilhe93} we also include surface tension here and neglect inertial effects.
Usually, the dynamics of this instability, either driven by evaporation and condensation or by surface diffusion, is slow, and therefore kinetic terms need not be considered here.
This is at least valid in the initial stage of the instability, but has to be corrected in the late regime when this omission leads to unphysical finite time singularities \cite{Kassner99,Brener03}.

\begin{figure}[th]
\begin{center}
\epsfig{file=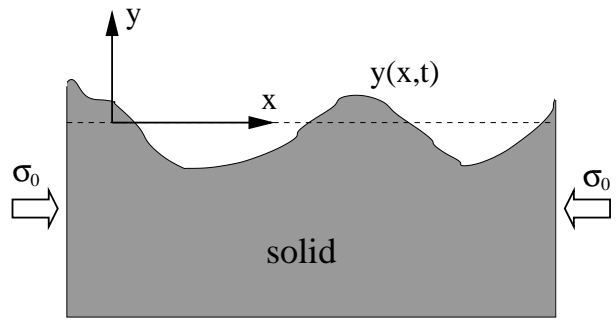, width=8cm}
\caption{The Grinfeld instability.}
\label{grin::fig1}
\end{center}
\end{figure}

The geometrical situation is sketched in Fig.~\ref{grin::fig1}.
The horizontally applied uniaxial loading $\sigma_{xx}^{(0)}=\sigma_{0}$ can be compressive or tensile.
Our goal is to calculate the stability of an originally flat surface.
Therefore we study the evolution of a slightly perturbed interface,
\[
y(x, t) = \Delta \cos kx.
\]
Here the boundary conditions are according to Eqs. (\ref{mech::eq3}, \ref{mech::eq4}) $\sigma_{nn}=\beta\kappa$ and $\sigma_{n\tau}=0$.
The latter one implies $u_{n\tau}=0$ along the interface.
This simplifies the general expression for the chemical potential, and we obtain
\[
\mu = \Omega \left( \frac{1}{2} \sigma_{ik}u_{ik} - \alpha \kappa + \beta \left[ \frac{\partial u_{\tau\tau}}{\partial n} - u_{\tau\tau}\kappa \right] \right).
\]
The strategy is to solve the elastostatic problem for a slightly perturbed surface first, then to calculate the chemical potential, and finally to implement a dynamical process at this surface, driven by spatial variations of the chemical potential.

Curvature in Cartesian coordinates is given by
\[
\kappa= \frac{y''}{(1+{y'}^2)^{3/2}}.
\]
For a two-dimensional plane-strain situation with iso\-tro\-pic elasticity we have
\[
\sigma_{xz}=\sigma_{yz} = 0,\quad \sigma_{zz} = \nu(\sigma_{xx}+\sigma_{yy}).
\]
Strains are
\begin{eqnarray*}
u_{xx} &=& \frac{1}{E} \left( (1-\nu^{2})\sigma_{xx} - \nu(1+\nu)\sigma_{yy} \right), \\
u_{yy} &=& \frac{1}{E} \left( (1-\nu^{2})\sigma_{yy} - \nu(1+\nu)\sigma_{xx} \right), \\
u_{xy} &=& \frac{1+\nu}{E}\sigma_{xy},\\
u_{xz} &=& u_{yz} = u_{zz} = 0.
\end{eqnarray*}
We use an Airy function $U(x,y)$ to express stresses,
\[
\sigma_{xx}=\frac{\partial^{2} U}{\partial y^{2}}, \quad \sigma_{xy} = - \frac{\partial^{2} U}{\partial x \partial y}, \quad \sigma_{yy} = \frac{\partial^{2} U}{\partial x^{2}},
\]
fulfilling the biharmonic compatibility condition $\Delta^{2}U=0$ and the elastic boundary conditions $\sigma_{n\tau}=0$, $\sigma_{nn}=\beta\kappa$.
We make the ansatz \cite{Nozieres93}
\begin{equation}
\label{Airy}
U = \frac{\sigma_{0} y^{2}}{2} + (A_{1}y+B_{1})e^{ky} \cos kx + (A_{2}y+B_{2})e^{2ky} \cos 2kx,
\end{equation}
with $A_1, B_1$ and $A_2, B_2$ being of first and second order in the surface profile amplitude $\Delta$ respectively.
Then we obtain from the boundary conditions up to second order
\begin{eqnarray*}
A_{1} = -\Delta(\beta k + \sigma_{0}),&\quad& B_{1} = \beta\Delta, \\
A_{2} = \frac{1}{2} k \Delta^2 (\sigma_0 + \frac{1}{2} k\beta), &\quad& B_{2} = \frac{1}{4} \Delta^2 \sigma_{0}.
\end{eqnarray*}
The total chemical potential at the interface is therefore to first order in $\Delta$ 
\begin{eqnarray*}
\mu &=& \Omega \Bigg( \frac{(1-\nu^{2})\sigma_{0}^{2}}{2E} + k\Delta \cos kx \Big[ -2\sigma_{0}^{2}\frac{1-\nu^{2}}{E} + \alpha k \\
&& - k\frac{\beta(1-\nu^{2})}{E} (3\sigma_{0} + 2\beta k) \Big] \Bigg) + {\cal{O}}(\Delta^2).
\end{eqnarray*}
In the particular case of surface diffusion,
\begin{equation}
v_n = \frac{D}{\alpha\Omega} \frac{\partial^2 \mu}{\partial \tau^2},
\end{equation}
with a kinetic coefficient $D$, the amplitude evolution, $\Delta=\Delta_{0} \exp(\lambda t)$, leads up to linear order to the spectrum
\begin{equation} \label{beta::eq24}
\lambda = D k^{3} \left( 2\sigma_{0}^{2}\frac{1-\nu^{2}}{E\alpha} - k + k\frac{\beta(1-\nu^{2})}{E\alpha} (3\sigma_{0} + 2\beta k) \right),
\end{equation}
assuming $k\geq 0$.
Obviously, the surface tension favors the Grinfeld instability.
Apart from the Grinfeld length,
\begin{equation}
L_{G} = \frac{E\alpha}{2(1-\nu^{2})\sigma_{0}^{2}},
\end{equation}
we have now an additional lengthscale in the system,
\begin{equation}
L_{\beta} := \frac{\beta}{\sigma_{0}}.
\end{equation}
Consequently, the spectrum can be expressed as
\begin{eqnarray*}
\lambda L_{G}^{4}/D &=& (kL_{G})^{3} \Big( 1-(kL_{G}) + \frac{3}{2} (kL_{G}) \frac{L_{\beta}}{L_{G}} \\
&&+ (kL_{G})^{2} \frac{L_{\beta}^{2}}{L_{G}^{2}} \Big).
\end{eqnarray*}
Typically, $L_{\beta}\ll L_{G}$, since $\sigma_{0}\ll E$, and thus the surface tension correction can be neglected.
However, the self-stress due to surface tension in Eq.~(\ref{beta::eq24}) leads to an instability even without external loading $\sigma_{0}$.

\section{Energy calculation}
\label{Grilhe}

In addition to the local chemical potential, also the total change of energy due to the morphological deformation of the surface can be calculated, leading to the same result.

We compute the energy which is contained in a strip of the width $2\pi/k$.
By symmetry, only even order contributions in $\Delta$ can appear.
The total elastic energy,
\[
E_{b} = \int\limits_{x=x_0}^{x_0+2\pi/k} \int\limits_{y=-\infty}^{y(x)} U_{b}\, dy\, dx
\]
is infinite due to the zeroth order contribution (homogeneous strain).
We introduced a shift $x_0$ which is irrelevant here since the energy content is the same in all strips of width of one period.
The second order contribution in $\Delta$ is
\[
E_{b}^{(2)} = - \frac{\pi \Delta^2}{E} (1-\nu^2)(\sigma_{0}^2 - k^2\beta^2).
\]
The surface energy,
\[
E_{\alpha} = \int\limits_{x=x_0}^{x_0+2\pi/k} \alpha \sqrt{1+{y'}^2}\, dx,
\]
is to second order
\[
E_{\alpha}^{(2)} = \frac{\alpha k \pi \Delta^2}{2}.
\]
Finally, the surface tension contribution
\begin{equation} \label{energy::eq1}
E_{\beta} = \int\limits_{x=x_0}^{x_0+2\pi/k} \beta u_{\tau\tau} \sqrt{1+{y'}^2}\, dx
\end{equation}
gives
\[
E_{\beta}^{(2)} = - \frac{\beta k\pi \Delta^2}{2E} (1-\nu^2)(4k\beta + 3\sigma_0).
\]
The expression (\ref{energy::eq1}) can be presented in a more suggestive form by the means of Eq.~(\ref{mech::eq1}),
\begin{eqnarray}
U_{\beta} &=& \beta \left[u_{\tau}(x_0+2\pi/k, y(x_0+2\pi/k)) - u_{\tau}(x_0, y(x_0))\right] \nonumber \\
&& - \int\limits_{x=x_0}^{x_0+2\pi/k} \beta\kappa u_n\, d\tau. \label{energy::eq2}
\end{eqnarray}
The first contribution vanishes for properly chosen offset $x_0=0$.
Also, it disappears for a closed boundary without corners.
The second contribution is the work done by the total displacement against the surface tension $\beta\kappa$.

Altogether, $E^{(2)} = E_{b}^{(2)} + E_{\alpha}^{(2)} + E_{\beta}^{(2)}$ results in
\begin{eqnarray}
E^{(2)} &=& \frac{\pi\Delta^2}{2}  \Big( -2\sigma_0^2 \frac{1-\nu^2}{E} + \alpha k \nonumber \\
&& - k \frac{\beta(1-\nu^2)}{E} (3\sigma_0 + 2k\beta) \Big).
\end{eqnarray}
The energy has the same structure as the spectrum (\ref{beta::eq24}), therefore the prediction of the threshold of instability is the same.
We point out that the elastic constants appear only in the combination $E/(1-\nu^2)$, since the scenario described here is plane strain.

At a first glance, the description of the Grinfeld instability using the chemical potential and the energy calculation seem to differ in a very important sense:
The chemical potential is first order in the perturbation amplitude $\Delta$;
in particular, it is sufficient to calculate all elastic fields only up to first order.
The use of energetic arguments always requires the calculation of second order terms.
In principle, it needs the elastic fields also up to second order, because for example in the bulk energy, also terms like $\sigma_{ij}^{(0)} u_{ij}^{(2)}$ appear.
This seems to contradict the former results using the chemical potential, which should contain the same physical information.
In fact, it turns out that the second order terms in the elastic fields do not enter into the final energy terms;
in other words, the same results can be obtained, setting $A_2=B_2=0$ in the Airy function above.
The explanation is now rather simple for the surface tension integral:
In the first representation (\ref{energy::eq1}), it seems that $u_{\tau\tau}$ has to be evaluated up to second order.
In the second representation (\ref{energy::eq2}), assuming that the first term vanishes, the curvature is already of first order in $\Delta$, and hence only the first order of $u_n$ is required.
For the elastic bulk energy it is known that in the present case of linear elasticity there is no interaction between the stress caused by externally applied forces and internal locked-in stress by surface tension \cite{Hirth82}, and therefore only first order terms are needed here.

Similar calculations of the total energy of a wavy surface were done in \cite{Grilhe93}.
However, the displacement fields without surface tension, as expressed by Eq.~(5) there, are incorrect;
the correct expressions can in principle easily be obtained from the Airy function (\ref{Airy}).
In particular, it would be necessary to include also third order harmonics $\cos 3kx$ to solve the elastic problem to the third order.
Furthermore, the vertical displacement becomes correctly to first order (without surface tension)
\[
u_y = -\frac{1-\nu^2}{E} \sigma_0 \Delta \cos kx + {\cal{O}}(\Delta^2),
\]
in contrast to Eq.~(5) in \cite{Grilhe93}.
Additionally, in the expression for the total energy of a system with surface tension, the work done by the surface tension against the (zeroth order) homogeneous displacement field was overlooked in \cite{Grilhe93}.
For a closed contour, this energy contribution can be written as
\begin{equation}
U_\beta = \int\limits_{\partial V} \beta\kappa u_n\, d\tau
\end{equation}
with the {\em total} normal displacement $u_n$.
For a non-closed surface also the corner contributions in Eq.~(\ref{energy::eq2}) have to be taken into account.

This underlines the superiority of our formal approach to derive the total energy and the chemical potential, because intuitive guesses for the work done for example by surface tension are not required.


\begin{acknowledgments}
I appreciate many useful discussions with Efim Brener.

This work has been supported by the Deutsche For\-schungs\-ge\-mein\-schaft (Grant No. SPP 1120).
\end{acknowledgments}

\appendix

\section{Conservation of momentum and angular momentum}
\label{conservation}

Here we derive the basic conservation laws which result from continuous symmetries of the Lagrangian (\ref{Lag::eq1}).
We perform a variation of the fundamental field $u_{i}$, assuming prescribed evolution of the volume $V(t)$ and $\partial V(t)$.
This means that a reshuffling of material by other processes than elastic displacement is possible;
the total number of atoms does not need to be conserved.
For simplicity, we assume the mass density $\rho$ to be constant and no external forces $\sigma_{in}^{(0)}$ to be present.

Using the expressions (\ref{Lag::eq1})-(\ref{Lag::eq4}) and (\ref{mech::eq5}), one gets
\begin{eqnarray}
\delta L &=& \int\limits_{V(t)} \left( \rho\dot{u}_{i}\delta\dot{u}_{i} - \sigma_{ik}\delta u_{ik} \right)dV + \int\limits_{\delta V(t)} \beta\kappa \delta u_{n}\, d\tau \nonumber \\
&=& \int\limits_{V(t)} \left( \rho \dot{u}_{i}\delta \dot{u}_{i} + \frac{\partial\sigma_{ik}}{\partial x_{k}}\delta u_{i} \right) dV \nonumber \\
&& + \int\limits_{\partial V(t)} \left( \beta\kappa \delta u_{n} - \sigma_{in}\delta u_{i} \right)\, d\tau \nonumber \\
&=& \frac{d}{dt} \int\limits_{V(t)} \rho\dot{u}_{i}\delta u_{i}\, dV, \label{conserv::eq1}
\end{eqnarray}
where the last step follows from the bulk equation (\ref{mech::eq2}) and boundary conditions (\ref{mech::eq3}, \ref{mech::eq4}).

Obviously, the Lagrangian (\ref{Lag::eq1}) is invariant with respect to a rigid body translation, $\delta u_{i} = \varepsilon_{i} = const.$, since $\delta\dot{u}_{i}=0$ and $\delta u_{ik}=0$.
On the other hand, by Eq.~(\ref{conserv::eq1}), $\delta L = \varepsilon_{i} dP_{i}/dt$, with the total momentum
\begin{equation} \label{conserv::eq2}
P_{i} = \int\limits_{V(t)} \rho \dot{u}_{i}\, dV.
\end{equation}
Therefore, the total momentum is conserved and the total surface tension force is zero.

Similarly, we can consider rigid body rotations, $\delta u_{i} = \epsilon_{ijk} \varepsilon_{j} x_{k}$, with a constant axial vector $\varepsilon_{i}$; $\epsilon_{ijk}$ is the totally antisymmetric tensor.
These rotations do not affect strain, $\delta u_{ik}=0$, and therefore again the Lagrangian (\ref{Lag::eq1}) is invariant, $\delta L=0$, since also $\delta\dot{u}_{i}=0$.
By (\ref{conserv::eq1}), we have $\delta L = \varepsilon_{j} d\ell_{j}/dt$ with the angular momentum
\begin{equation}
\ell_{i} = \int\limits_{V(t)} \rho \epsilon_{ijk} x_{j} \dot{u}_{k}\, dV.
\end{equation}
Thus the total angular momentum is conserved, too.

Notice that both conservation laws are not trivial, since the body can change its size and therefore its number of atoms:
the normal velocity describes the local extension or shrinkage of the body, and is not specified here.

An example is a solid droplet in its melt phase, for example in a temperature gradient.
By crystallization the body can grow at one part of the interface and recede at another;
this means that the droplet can wander, though all its ``solid'' atoms are at rest.
In other words, the body moves, but its total momentum (\ref{conserv::eq2}) vanishes, since the displacement is constant.

\section{An integral theorem}
\label{deriv}

We look at an energy functional of the form
\[
W[y(x,t), t] = \int\limits_{V(t)} f\, d\tau = \int\limits_{x_{0}}^{x_{1}} f(x, y, y', t) \sqrt{1+{y'}^{2}}\, dx
\]
in Cartesian coordinates.
Here the prime denotes the derivative with respect to $x$.
We consider now the rate of change of the energy $W$:
During a timestep $\delta t$ both the function value $f$ and the interface position $y$ change.
\begin{figure}[th]
\begin{center}
\epsfig{file=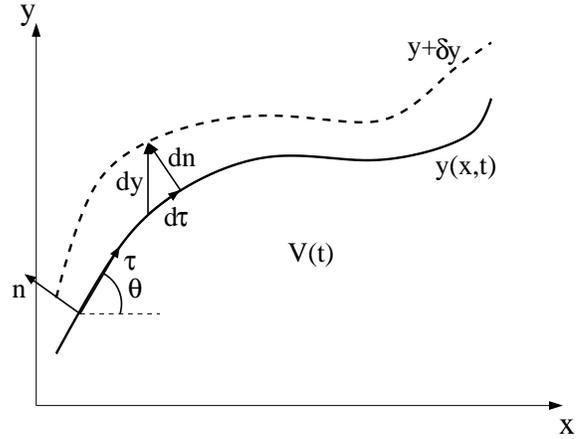, width=7.5cm}
\caption{The interface at time $t$ is depicted by the solid curve, the shifted interface at $t+\delta t$ by the dashed curve.}
\label{theo::fig1}
\end{center}
\end{figure}
Using the notations
\[
\delta y = \frac{\partial y}{\partial t}\, \delta t,\quad
\delta y' = \frac{\partial^{2} y}{\partial x \partial t}\, \delta t,
\]
the rate of change is to first order in $\delta t$
\begin{widetext}
\begin{eqnarray*}
\delta W &=& W[y(x,t+\delta t), t+\delta t] - W[y(x,t), t] \\
&=& \int\limits_{x_{0}}^{x_{1}} \Bigg\{ f \frac{y'}{(1+{y'}^{2})^{1/2}} \delta y' + (1+y'^{2})^{1/2} \left[ \frac{\partial f}{\partial y} \delta y 
+ \frac{\partial f}{\partial y'} \delta y'
+ \frac{\partial f}{\partial t} \delta t \right] \Bigg\} dx \\
&=& \int\limits_{x_{0}}^{x_{1}} \Bigg\{ -\left[ f \frac{y'}{(1+{y'}^{2})^{1/2}} \right]' \delta y 
+  \left[ \frac{\partial f}{\partial y} \delta y + \frac{\partial f}{\partial t} \delta t \right] (1+y'^{2})^{1/2}  
- \left[ \frac{\partial f}{\partial y'} (1+y'^{2})^{1/2} \right]' \delta y \Bigg\},
\end{eqnarray*}
where we performed an integration by parts, assuming that the boundary contributions vanish;
this is automatically fulfilled for a closed contour. 
The prime denotes the total spatial derivative (with fixed time $t$), i.e.
\[
[\ldots]' = \frac{\partial [\ldots]}{\partial x} + \frac{\partial [\ldots]}{\partial y} y' + \frac{\partial [\ldots]}{\partial y'} y''.
\]
Therefore 
\begin{eqnarray*}
\delta W &=& \int\limits_{x_{0}}^{x_{1}} \Bigg\{ - \Bigg[ \frac{\partial f}{\partial x} + \frac{\partial f}{\partial y} y' 
+ \frac{\partial f}{\partial y'} y''\Bigg] \frac{y'}{(1+{y'}^{2})^{1/2}} \delta y 
- f\left[\frac{y''}{(1+{y'}^{2})^{1/2}} - \frac{{y'}^{2}y''}{(1+{y'}^{2})^{3/2}} \right] \delta y 
+ \frac{\partial f}{\partial y} (1+{y'}^{2})^{1/2} \delta y \\
&& - \left[ \frac{\partial^{2} f}{\partial x\partial y'} + \frac{\partial^{2} f}{\partial y\partial y'} y' + \frac{\partial^{2} f}{\partial y'^{2}} y''\right] (1+{y'}^{2})^{1/2} \delta y
- \frac{\partial f}{\partial y'} \frac{y'y''}{(1+{y'}^{2})^{1/2}} \delta y + \frac{\partial f}{\partial t} (1+{y'}^{2})^{1/2} \delta t \Bigg\} dx.
\end{eqnarray*}
\end{widetext}
Curvature is given by
\[
\kappa = \frac{y''}{(1+y'^{2})^{3/2}},
\]
and we have normal and tangential vectors
\[
\bm{n} = \frac{1}{(1+y'^{2})^{1/2}} \left(\begin{array}{c} -y' \\ 1 \end{array}\right), \quad
\bm{\tau} = \frac{1}{(1+y'^{2})^{1/2}} \left(\begin{array}{c} 1 \\ y' \end{array}\right).
\]
Instead of Cartesian coordinates $(x, y, y')$ (where $y'$ is used to describe the slope), we use now local coordinates $(n, \tau, \theta)$.
$n$ is the normal, $\tau$ the tangential coordinate and $\theta$ the inclination angle of the surface.
With $\theta=\arctan y'$,
we get the following transformation rules:
\begin{eqnarray*}
\frac{\partial}{\partial\theta} &=& (1+y'^{2}) \frac{\partial}{\partial y'},  \\
\frac{\partial}{\partial n} &=& \frac{-y'}{(1+y'^{2})^{1/2}} \frac{\partial}{\partial x} 
+ \frac{1}{(1+y'^{2})^{1/2}} \frac{\partial}{\partial y}, \\
\frac{\partial}{\partial \tau} &=& \frac{1}{(1+y'^{2})^{1/2}} \frac{\partial}{\partial x}
+ \frac{y'}{(1+y'^{2})^{1/2}} \frac{\partial}{\partial y}.
\end{eqnarray*}
Comparing this with the above expression gives
\begin{eqnarray*}
\delta W &=& \int\limits_{x_{0}}^{x_{1}} \Bigg\{ \Bigg[ \frac{\partial f}{\partial n} - f\cdot \kappa - \frac{\partial^{2}f}{\partial\tau\partial\theta} - \kappa \frac{\partial^{2} f}{\partial\theta^{2}} \Bigg] \delta y \\
&& + \frac{\partial f}{\partial t} (1+{y'}^{2})^{1/2} \delta t \Bigg\}  dx.
\end{eqnarray*}
The next step is to show that a vertical shift $\delta y$ is related to a normal shift $\delta n$ according to
\begin{equation} \label{theo::eq2}
\delta y\, dx = \delta n\, d\tau = v_{n} \delta t\, d\tau,
\end{equation}
where the last relation defines the interface normal velocity $v_{n}$.
By substitution
\begin{equation} \label{theo::eq1}
\delta y dx = \delta y \frac{dx}{d\tau}d\tau
\end{equation}
with the arclength element $d\tau^{2}=dx^{2}+dy^{2}$ (see Fig.~\ref{theo::fig1}).
Normal, tangential and vertical displacements are
\[
\delta\bm{\tau} = \left(
\begin{array}{c}
\delta\tau_{x} \\
\delta\tau_{y}
\end{array}
\right),
\qquad
\delta \bm{n} = \left(
\begin{array}{c}
\delta n_{x} \\
\delta n_{y}
\end{array}
\right),
\qquad
\delta \bm{y} = \left(
\begin{array}{c}
0 \\
\delta y
\end{array}
\right)
\]
with $\delta \bm{n} + \delta\bm{\tau} = \delta \bm{y}$, thus
\[
\delta\bm{\tau} = \left(
\begin{array}{c}
-\delta n_{x} \\
\delta y - \delta n_{y}
\end{array}
\right).
\]
Orthogonality of normal and tangential variation requires
\[
\delta n^{2} = \delta n_{x}^{2} + \delta n_{y}^{2} = \delta y\, \delta n_{y}.
\]
Furthermore, the normal vector is perpendicular to the interface $(dx, dy)$, and therefore
\[
\delta n_{y} = \frac{dx}{d\tau} \delta n.
\]
Combining the preceding expressions gives
\[
\delta n = \frac{dx}{d\tau}\delta y,
\]
which, together with (\ref{theo::eq1}), proves the statement (\ref{theo::eq2}) above.

Using this result, we finally obtain
\begin{equation} \label{deriv::eq2}
\frac{dW}{dt} = \int\limits_{V(t)} \Bigg\{ v_{n} \Bigg[ \frac{\partial f}{\partial n} - f\cdot \kappa - \frac{\partial^{2}f}{\partial\tau\partial\theta} 
- \kappa \frac{\partial^{2} f}{\partial\theta^{2}} \Bigg]  + \frac{\partial f}{\partial t} \Bigg\}\, d\tau . 
\end{equation}
The expression in brackets reflects the change of energy due to motion of the interface;
in particular, it consists of a contribution caused by the motion of the interface in normal direction, the elongation of a curved length element and a tangential translational and rotational contribution;
the last term is due to the change of the function $f$ in time.

Similarly, we can derive the functional derivative $\delta W/\delta n$, now at a fixed time $t$.
It corresponds to $v_{n}=\delta(x_{n})$, and thus
\begin{equation} \label{deriv::eq1}
\frac{\delta W}{\delta n} = \frac{\partial f}{\partial n} - f\cdot \kappa - \frac{\partial^{2}f}{\partial\tau\partial\theta} - \kappa \frac{\partial^{2} f}{\partial\theta^{2}}.
\end{equation}


\end{document}